\documentstyle[epsfig,aaspp4]{article}
\begin{document}
\title{Cosmic Ray Protons and Magnetic Fields in Clusters of
Galaxies\\ and their Cosmological Consequences} \author{Torsten
A. En{\ss}lin, Peter L. Biermann} \affil{Max-Planck-Institut f\"{u}r
Radioastronomie, Auf dem H\"{u}gel 69, D-53121 Bonn, Germany}
\author{Philipp P. Kronberg}
\affil{Department of Astronomy, University of Toronto, 60, St. George
Street, Toronto, Canada \nolinebreak M5S \nolinebreak 1A7} \and
\author{Xiang-Ping Wu} \affil{Beijing Astronomical Observatory, Chinese
Academy of Science, Beijing 100080, China}
\begin{abstract}
The masses of clusters of galaxies estimated by gravitational lensing
exceed in many cases the mass estimates based on hydrostatic
equilibrium.  This may suggest the existence of nonthermal
pressure. We ask if radio galaxies can heat and support the cluster
gas with injected cosmic ray protons and magnetic field densities,
which are permitted by Faraday rotation and gamma ray observations of
clusters of galaxies. We conclude that they are powerful enough to do
this {  within a cluster radius of roughly 1 Mpc }. If
present, nonthermal pressures could lead to a revised estimate of the
ratio of baryonic mass to total mass, and the apparent baryonic
overdensity in clusters would disappear.  In consequence, $\Omega_{\rm
cold}$, the clumping part of the cosmological density $\Omega_{o}$,
would be larger than $0.4\,h_{50}^{-1/2}$.
\end{abstract}
\section{Introduction}

Clusters of galaxies, being the most massive coherent objects in the
universe, are important probes of the cosmological density. X-ray
observations allow a measurement of the density profile of the hot
intracluster gas, which dominates the visible, baryonic mass $M_{\rm
b}$ of a cluster. Estimates of the total mass $M_{\rm tot}$ can be
made from a virial analysis of the galaxy velocities, by integration
of the hydrostatic equilibrium between gravitational forces and
thermal pressure, or by analyzing background objects that are
gravitationally lensed by the cluster.  Calibrating $\Omega_{\rm b}$
from the standard Big Bang nucleosynthesis with the value $\Omega_{\rm
b}/\Omega_{\rm cold} \approx M_{\rm b}/M_{\rm tot}$ given by clusters
should give then a lower limit to $\Omega_{o}$, if $\Omega_{\rm cold}$
is the part of the matter which clumps on the scale of clusters.
Unfortunately, the mass derived from velocity dispersion and from
hydrostatic equilibrium appears to be often much lower than the mass
derived by lensing methods.

Miralda-Escud\'{e} \& Babul (1995) \nocite{miralda95} derived the mass
of Abell 2218 and Abell 1689 from gravitationally lensed arcs and from
X-ray observations and found a mass shortfall of a factor 2.5 $\pm$
0.5.  Similar work done by Wu (1994) \nocite{wu94} gave a factor of 3
-- 6 within a central radius of ~300 kpc $h_{50}^{-1}$ for four
different clusters. Even in the rich, early cluster RXJ1347.5-1145 at
$z=0.451$ a mass discrepancy of a factor 2 -- 3 is reported by
Schindler et al. (1996).  Conversely, Squires et al. (1995)
\nocite{squires95} have  re-examined Abell 2218 with HST
images and found accordance between the different mass determinations
at a radius of 800 $h_{50}^{-1}$ kpc, applying a weak lensing method
that reconstructs the mass distribution by using the distortion of
background galaxies.  However, they note that their lensing mass could
be too low, since it depends on assumptions about the mass
distribution outside the image radius. But if their mass values are
correct, then the mass discrepancy would be largely removed for
regions far outside the core.  Allen, Fabian, \& Kneib (1995)
\nocite{allen95} show for the cluster PKS0745 that a two-temperature
model of the gas, consistent with a strong cooling flow, has no discrepancy
between the hydrostatic mass and an arc-determined mass at a small
radius of $46 \, h_{50}^{-1}$ kpc.  D.-W. Kim \& Fabbiano (1995)
\nocite{kim95} find a discrepancy between masses estimated from X-rays
and the virial masses from the velocity dispersions of the galaxies in
the NGC 507 group.  But while many measurements give virial masses as
low as the X-ray masses (Bahcall \& Lubin 1993) \nocite{bahcall93},
these virial mass estimates could be a strong underestimate of the
total masses of clusters, predicted from the appearance of dynamical
friction in simulations of the galaxy and dark matter content of
clusters (Serna, Alimi, \& Scholl 1994; Carlberg 1994a)
\nocite{serna94,carlberg94a}. This would be consistent with a number
of observations of velocity and luminosity segregation of galaxies in
clusters (Yeppes, Dom\'{i}nguez-Tenreiro, \& Del Pozo-Sanz 1991;
Biviano et al.  1992; Buote \& Canizares 1992; Carlberg 1994b; Loveday
et al. 1995)
\nocite{yeppes91,biviano92,buote92,carlberg94b,loveday95}.  The masses
of clusters are overestimated if the measured velocity dispersion is
seriously affected by infalling galaxies. But a parameter-free
examination of the Coma cluster with 1500 galaxy positions and 450
measured velocities by Merritt \& Gebhardt (1996) \nocite{merritt96}
shows that, even in this case with solid statistics, the total mass of
the cluster is poorly defined and could be several times the value
derived by assuming that mass follows light.  The average trend of a
large sample of clusters show a clear signal for a strong mass
discrepancy between lensing and hydrostatic masses, which extends up
to a radius of 1 Mpc (Wu \& Fang 1996).

Several possibilities have been suggested to resolve the mass
discrepancy between X-ray and lensing masses, including a projection
effect of an asymmetrical matter distribution (for a discussion see
Miralda-Escud\'{e} \& Babul 1995) \nocite{miralda95}.  Additionally,
substructuring can cause significant uncertainties in the computation
of the dynamical cluster mass. Loeb \& Mao (1994) \nocite{loeb94} and
Steigman \& Felten (1995) \nocite{felten95} explained the mass
discrepancy as due to ignorance of nonthermal pressure in the
hydrostatic equilibrium of the intracluster medium (ICM). 

In this paper we want to focus on the role that the nonthermal cosmic
ray pressure could play in supporting the intracluster ionized gas.
In the disk of our Galaxy cosmic rays (CR) are trapped in the galactic
magnetic field, which is frozen into the interstellar gas through its
ionized component. Recent evidence that the ICM of galaxy clusters is
permeated by significant magnetic fields suggests that a similar
trapping of CR by the ICM field occurs, although direct measurement of
the ICM cosmic ray component is more difficult to make.  For our
galaxy, the effects of the cosmic ray gas component are reasonably
well understood --- {\it cf.}  Parker's (1969) review article
\nocite{parker69}. If the pressure of these nonthermal phases exceeds
the thermal pressure by a factor of order unity, instabilities will
grow, and convective processes will result. This sets a maximum level
at which nonthermal constituents will likely be present.  If there are
a sufficient number of powerful sources for these pressures, then the
nonthermal pressures should approach roughly this limiting value ---
which we can also consider as a {\it maximum} value. Since we might
expect a constant factor between thermal and CR-pressure, we
parametrize the CR-energy density as a fraction of the thermal energy
\begin{equation}
\label{ECR}
\varepsilon_{\rm CR}(r) = \alpha_{\rm CR} \, \varepsilon_{\rm th}(r) \, .
\end{equation}

For our Galaxy, Parker showed that a magnetic plus cosmic ray pressure
of 1.5 times the thermal pressure results from the following
measurements and basic physical considerations:
\begin{enumerate}
\begin{enumerate}
\item
Direct observation of the cosmic ray spectrum in the solar
neighborhood, combined with observation of Faraday rotation, which
quantitatively establishes both cosmic ray and magnetic pressure
component in the interstellar medium.
\item
Hydrostatic effects, which determine the (observable) scale height of
the gas distribution in our galaxy, which is sensitive to gravity and
pressure.
\item
A theoretical limit to the gas height scale, given by pressure
limiting instabilities for magnetic fields perpendicular to the
gravitational forces in a disk filled with a cosmic ray gas.
\end{enumerate}
\end{enumerate}

A key aim of this paper is to adapt our most recent knowledge of the
multi-phase galactic disk gas to the intracluster medium of galaxy
clusters. By analogy to the interstellar medium, we propose that the
ICM is kept somewhere close to a ``stability limit'' by sources of
cosmic rays and magnetic fields. Whereas the cosmic ray component of
the ICM comes from supernova remnants, we propose that radio galaxies
are the corresponding CR sources in galaxy clusters. 
 We note the detailed analogy to the interstellar medium: Supernova
 remnants provide the interstellar medium with CRs and high velocity
 motion in a phase of their evolution {\it after} they are clearly
 discernible as radio and/or X-ray sources. Similary, we expect radio
 galaxies to distribute most of their cosmic rays, magnetic fields and high
 velocity motion in an evolutionary phase {\it after} they are
 recognizable in the radio.
Thus, we require a basis for assuming that there are sufficient CR
sources to affect the nonthermal pressure component in the ICM.
Although the field topology in clusters is not yet well understood, we
can estimate the pressure-determining factor by using approach (b) to
get a rough estimate.

We investigate the question of a new pressure component in clusters
due to cosmic rays from three different lines of investigation and
argument.

(i) We will argue from current and recent observations that magnetic
fields and energetic protons provide a substantial internal energy
component in clusters of galaxies. Further, there is hope that these
CR protons should be detectable in the near future, if they are indeed
present. The main component of cosmic rays and magnetic fields in the ICM
should come from radio galaxies (RG), which are frequently found in
the central regions of clusters. Over the lifetime of galaxy clusters,
radio galaxies should have injected fields and energetic particles,
whose combined energy is at least equal to the present-epoch thermal
energy content of the central region.  To demonstrate this, we
calculate the energy output from clusters by X-ray cooling and their
thermal energy-content. We compare these numbers with the jet
power-input from radio galaxies into the ICM at the present cluster
epoch, and integrated over the clusters' history. We do this
global calculation for clusters and radio galaxies by integrating,
separately, their X-ray- and radio-luminosity functions, and by using
statistical correlations between properties and luminosities of these
objects.

(ii) Nonthermal pressure should show up in a comparison of gravity
forces and thermal pressure gradients. The latter are used in deriving
hydrostatic masses of clusters, and it is these masses which are lower
by a factor of typically 3 than masses derived by gravitational
lensing.  The missing pressure component that is required to ``close''
the discrepancy is approximately 2 times the thermal pressure. 

(iii) Due to the expected complicated field topology and local
variations of field strength, a universal, stable and time-independent
configuration of fields, cosmic-rays and thermal gas is
unlikely. However, turbulent transport of magnetic fields and cosmic
rays over large distances in a cluster should permit a factor between
nonthermal and thermal pressure that is comparable to the one in our
(much smaller) galactic disk.

In the following sections we present arguments to show that, under
reasonable assumptions, the energy output from radio galaxies {\it
can} provide energy in the required order of magnitude to close the
gap between hydrostatic and direct gravitational estimates of
masses { at least in the central regions of galaxy
clusters}. Throughout this paper we adopt $H_{o}=50 \, h_{50}$ km
s$^{-1}$ Mpc$^{-1} $ and $q_{o}=0.5$. The term cosmic rays (CRs) will
primarily stand for energetic protons.
\section{The energy content of the ICM}
\subsection{The thermal energy}
The cosmological density of the X-ray luminosity of galaxy clusters
can be calculated by an integration of the X-ray luminosity function
of clusters of galaxies, proposed by Edge et al. (1990) \nocite{edge90}
in the 2 -- 10 keV band
\begin{equation}
\label{eq.XLF}
\frac{dN}{dL_{X,2-10}} = A L_{X,2-10}^{-\alpha} \exp(- L_{X,2-10}/L_o
) \, {\rm Mpc^{-3}},
\end{equation}
where the luminosity $L_{X,2-10}$ is given in units of $10^{44}\, {\rm
erg\,s^{-1}}$ and $A=10^{-6.57\pm 0.12}$, $\alpha = 1.65 \pm 0.26$ and
$L_o = 11_{-0.2}^{+0.4}$ ($cl = 90\%$ within this section). We can use
the strong correlation $L_{X} = 10_{}^{B_{\rm bol}}\,
L_{X,2-10}^{A_{\rm bol}}$ ($A_{\rm bol}= 0.94 \pm 0.03, B_{\rm bol}=
3.02 \pm 0.15$ and the luminosities in ${\rm erg \, s^{-1}}$) (Edge \&
Steward 1991) \nocite{edge91} to obtain an estimate of the bolometric
luminosity of clusters. Integrating from the lower limit $10^{43}\,
{\rm erg\, s^{-1}}$ to infinity gives $L_{X} \approx 2 \cdot 10^{40}
\, h_{50} \, \mbox{W Gpc}^{-3}$.
In order to estimate the thermal energy content we use a
$\beta$-model for the cluster gas
\begin{equation}
n_{\rm e}(r)= n_{o} \, \left( 1+(r/r_{\rm core} )^{2} \right)^{-3/2
\beta}
\end{equation}
and the cooling function for thermal bremsstrahlung
\begin{equation}
\label{xray}
\Lambda_{X}(T) = \Lambda_{o} \, n_{\rm e}^{2}\,(k_B T)^{1/2} \, ,
\end{equation}
where $n_{\rm e}$ is the electron density, $T$ the temperature and
$\Lambda_o = 5.96 \cdot 10^{-24} {\rm erg \, s^{-1}\,cm^{3} \,
keV^{-1/2}}$. An integration of the luminosity to the largest observed
radius, $r_{\rm obs} \approx 10 \, r_{\rm core}$, gives 
\begin{equation}
\label{eq.lxint}
L_{X} = \Lambda_o \, n_o^2\, (k_B T)^{1/2}\, r_{\rm core}^3 \,
f(r_{\rm obs}/r_{core}, 3 \beta)\, ,
\end{equation}
where 
\begin{equation}
f(x, \gamma) = 4 \pi \, \int\limits_{0}^{x} \, dy\,
\frac{y^2}{(1+y^2)^{\gamma}} = 2 \pi \, B_{\frac{x^2}{1+x^2}}( 
\frac{3}{2}, \gamma - \frac{3}{2})
\end{equation}
depends only on the geometry. $B$ is the incomplete Beta function. The
thermal energy content ($3\, n\,k_{B}\,T$) integrated out to a radius
$r$ yields, with equation (\ref{eq.lxint}),
\begin{equation}
E_{\rm th}(r) = 3\,(k_B T)^{3/4}\, L_{X, bol}^{1/2} \,r_{\rm core}^{3/2} \,
\frac{f(r/r_{\rm core}, 3\beta/2)}{\sqrt{f(r_{\rm obs}/r_{\rm core}, 3
\beta)}} .
\end{equation}
Using the good correlation $k_B T = 10_{}^{B_{\rm T}}
L_{X,2-10}^{A_{\rm T}}$ ($A_{\rm T} = 0.28 \pm 0.05, B_{\rm T} =
-11.73 \pm 0.20, k_B T$ in keV and $L_{X,2-10}$ in ${\rm erg \,
s^{-1}}$) (Edge \& Steward 1991) \nocite{edge91} we can translate
temperature into luminosity with good reliability.  Given that the
correlation of $r_{\rm core}$ with $T$ is weak, we assume a constant
characteristic core radius of $r_{\rm core} = 250 \, h_{50}^{-1}$ kpc
and $\beta = 0.6$ for all clusters.  Integration of the luminosity
function (\ref{eq.XLF}) gives an averaged thermal energy of the ICM
within the cooling radius $r_{\rm cool} \approx 1.5 \, r_{\rm core}$
which is $E_{\rm th}(r_{\rm cool}) \approx 3 \cdot 10^{57} \,
h_{50}^{1/2} \,\mbox{J Gpc}^{-3}$. This number is nearly independent
of $\beta$ within the range 0.6 ... 1.0 and of $r_{\rm obs}$ as long
as it is several times $r_{\rm core}$. The thermal energy within
$1.5 \,$Mpc$\,h_{50}^{-1}$ depends even more on $\beta$ and is one
order of magnitude higher. Shifting $\beta$ to 1.0 would lower the
thermal energy content within $1.5 \,$Mpc$\,h_{50}^{-1}$ by 40\%. 

\subsection{Recent determinations of intracluster field strengths}

Statistical evidence for widespread intracluster magnetic fields comes
from a combination of Faraday rotation measurements, combined with
X-ray determined electron densities in the same ICM gas which produces
the Faraday rotation. The most recent analysis, by K.-T. Kim, Tribble,
\& Kronberg (1991) \nocite{kim91} for a sample of 50 Abell clusters,
indicates a typical ICM field strength of $\sim 2 \mu \mbox{G} (l_{\rm
frs}/10 \,{\rm kpc})^{-1/2}h_{50}^{-1/2}$ (where $l_{\rm frs}$ is the
typical field reversal scale). This estimate, as (Kim et al. 1991)
\nocite{kim91} argue, is likely to increase, especially in the cores,
due to various uncertain factors such as smaller turbulence scales
({\it cf.} Feretti et al. 1995) \nocite{feretti95}, and the fact that
their average result (which used background radio sources) is weighted
toward sightlines outside of the cluster core region where the field
strength is likely weaker.   

It is interesting to note that, in some clusters, high field
values have been measured, and approximate equality between magnetic
and thermal pressure has been established ({\it cf.} Taylor and Perley
1993)\nocite{taylor93}.  These observations, of differential ``Faraday
screen'' effects over extended radio lobes of powerful radio galaxies
inside of clusters, reveal intracluster field strengths of order $30
\mu \mbox{G}\,h_{50}^{-1/2}$, {\it e.g.} in the core of the Hydra
cluster (David et al. 1990) \nocite{david90}. A similar result was
obtained for the host cluster of Cygnus A (Taylor \& Perley 1993)
\nocite{taylor93}.  It may not be a coincidence that the clusters with
powerful radiogalaxies also contain a cooling flow (Christodoulou \&
Sarazin 1996).

The recent trend of results for, {\it e.g.}, the Coma cluster is that
smaller field reversal scales have emerged as the resolution of the
Faraday RM images of cluster head-tail sources has improved. This
causes estimates of core ICM field strengths to increase
correspondingly ({\it cf.} Felten (1996) for a detailed discussion on
this point). Thus, the reversal scales established by new, higher
resolution images of the extended head-tail source in the Coma Cluster
give cluster core ICM field strengths of 5 $\mu \mbox{G}$ or higher
(Feretti et al. 1995).

Recent Faraday rotation data suggest that the general cluster 
population (as distinct from clusters presently known to have
cooling flows) are able to produce fields in their core regions 
(by whatever mechanism) of up to $10 \mu \mbox{G}\,h_{50}^{-1/2}$ 
or more.  This is consistent with the trend of recent attempts to 
measure cluster ICM magnetic field strengths (Kim {it et al.} 1990, 
Kim {it et al.} 1991, Feretti {\it et al.} 1995). 

If the turbulence scale, $l_{\rm frs}$, is
as small as $\sim 0.1$ kpc, then the sightline-averaged magnetic field
would as much as $\sim 20 \,\mu {\rm G}
\,h_{50}^{-1/2}$. The resulting magnetic pressure would, at that
level, be of the order of the thermal pressure. { Since the
strength of the fields and the typical field scale depend on the field
reversal scale, $l_{\rm frs}$ is kept in all formulae as a free
parameter -- this allows to use them in the case of weak ($l_{\rm frs}
\approx 10$ kpc) and strong ($l_{\rm frs} \approx 0.1$ kpc) magnetic
fields. }

Field values of this order support our notion of an additional 
non-thermal pressure component in cluster cores, although they 
need not be precisely in energy equipartition with the CR protons. 
It is also relevant to note that they are also comparable with 
(independently estimated) equipartition field values in weak 
relics of ``old'' extended extragalactic radio sources 
({\it cf.} Kronberg 1994) \nocite{kronberg94} which are near
$10\, h_{50}^{2/7}\,\mu\mbox{G}$ (Miley 1980) \nocite{miley80}. 

{ But there is no evidence as yet for or against dynamically
important fields outside of cluster core regions. }

\subsection{The gamma-ray emission and intracluster cosmic ray densities}
\placetable{tab.gammapred}
\begin{deluxetable}{lccc}
\tablenum{1}
\tablecaption{\label{tab.gammapred}Expected Gamma-Ray Fluxes}\
\tablehead{
\colhead{Cluster} &
\colhead{$F_{\gamma}$ this paper (counts ${\rm cm^{-1}\,s^{-1}}$)}  &
\colhead{$F_{\gamma}$ from Dar \& Shaviv (1995) \nocite{dar95}
(counts ${\rm cm^{-1}\,s^{-1}}$)} &
\colhead{$F_{\gamma}$ EGRET upper limit (counts ${\rm
cm^{-1}\,s^{-1}}$)} 
}

\tablehead{
\colhead{Cluster} &
\colhead{$F_{\gamma}$ this paper}  &
\colhead{$F_{\gamma}$ from Dar \& Shaviv (1995) \nocite{dar95}} &
\colhead{$F_{\gamma}$ EGRET upper limits}
\nl
& \colhead{(counts ${\rm cm^{-2}\,s^{-1}}$)} &
 \colhead{(counts ${\rm cm^{-2}\,s^{-1}}$)} &
\colhead{(counts ${\rm cm^{-2}\,s^{-1}}$)} }
\startdata
 A426 Perseus  & $12 \cdot 10^{-8}$& $10 \cdot 10^{-8}$&\nodata \nl
 Ophiuchus &   $9 \cdot 10^{-8}$ &  \nodata            &\nodata \nl
 A1656 Coma &  $6 \cdot 10^{-8}$&$5 \cdot 10^{-8}$&$4\cdot 10^{-8}$ \nl
 M87 Virgo & $3 \cdot 10^{-8}$& $22 \cdot 10^{-8}$&$4\cdot 10^{-8}$\nl
\enddata
\tablecomments{ Expected gamma-ray fluxes above 100 MeV from X-ray
luminous clusters. The first column gives the values calculated with
equation (\ref{xgamma}), data from (David et al. 1993)
\nocite{david93} and $\alpha_{\rm CR}=1$ and the second the values
given in (Dar \& Shaviv 1995) \nocite{dar95}. The last column gives
measured $2\,\sigma$ upper limits ($cl = 95\%$) from EGRET (Sreekumar
et al. 1996) \nocite{sreekumar96}}
\end{deluxetable}
Dar \& Shaviv (1995) \nocite{dar95} proposed to explain the diffuse
extragalactic gamma ray spectrum, and predict detectable gamma ray
fluxes from the nearest clusters. They assume that the cosmic ray
density in galaxy clusters is similar to that in our own galaxy and
that the density has no radial dependence. We use a more realistic
model, in which the cosmic ray energy $\varepsilon_{\rm CR}$ density
scales with the thermal energy density of the gas:
\begin{equation}
\varepsilon_{\rm CR}(r) = 3\, n_{\rm e}(r) \,k_B T\,\alpha_{\rm CR}
\end{equation}
where $\alpha_{\rm CR}$ is the scaling ratio between the thermal and
CR energy densities defined in equation (\ref{ECR}). We now estimate
the production rate for gamma rays above 100 MeV by $\pi_o$-decay
after hadronic interactions of the energetic protons with the
background gas. This is given by
\begin{equation}
\label{gammaray}
\frac{dn_{\gamma}(> 100{\rm MeV})}{dt} = q_{\gamma} \, \varepsilon_{\rm CR}\,
n_{\rm target} = 3\,q_{\gamma} \, n_{\rm e}^2 \,k_B T\,\alpha_{\rm CR}
\end{equation}
where $n_{\rm target}$ is the proton density, and the parameter
$q_{\gamma} = 0.39\cdot 10^{-13} {\rm cm^3\, erg^{-1} \,s^{-1}}$
applies to a proton spectrum similar in slope to that observed in our
galaxy ({\it cf.} Drury, Aharonian, \& V\"{o}lk 1994)
\nocite{drury94}.  We compare equation (\ref{gammaray}) with the
rate of thermal bremsstrahlung (equation (\ref{xray})) to obtain a
ratio for the observed fluxes gamma-rays $F_{\gamma}({\rm > 100 MeV})$
and X-rays $F_X$ for galaxy clusters
\begin{equation}
\label{xgamma}
\frac{F_{\gamma}({\rm > 100 MeV})}{F_X/{\rm erg}} = 32 \left(
\frac{k_B T}{\rm keV} \right)^{1/2} \alpha_{\rm CR} \, .
\end{equation}
We note that this ratio is practically independent of the assumed
cosmology. Taking temperatures and X-ray fluxes from David et
al. (1993) \nocite{david93} and assuming $\alpha_{\rm CR} = 1$, we
list in table \ref{tab.gammapred} our calculation of the expected
gamma-ray fluxes for some of the brightest clusters along with fluxes
calculated by Dar \& Shaviv (1996) \nocite{dar95}. The discrepancy
between our fluxes and theirs are due to the different CR-density
profiles. We note, that if the the CRs have a steeper radial
density profile than the gas -- reasonable due to the mostly central
injection and due to a decreasing mass discrepancy with radius -- our
gamma ray fluxes have to be reduced.

If there is an anticorrelation between the presence of cooling flows
and cosmic ray protons due to the possibility of a slowing down of the
cooling instability by nonthermal pressure, then our calculated
values for Perseus and Ophiuchus could be too high, since these have
strong cooling flows.

The above fluxes are close to the detection limit of EGRET, which
suggests that the unknown factor $\alpha_{\rm CR}$ between the CR-and
the thermal energy density in these clusters could be determined in
the near future for some clusters. For the Coma and Virgo clusters
upper limits from EGRET measurements are now available and given in
table \ref{tab.gammapred}. In the case of Coma, the upper limit restricts the
unknown parameter $\alpha_{\rm CR}$ to be lower than $2/3$ or the CRs
are in equipartition only within the central region. Unfortunately the
expected fluxes above some TeV are orders of magnitude too low to be
detectable by the HEGRA or other airshower experiments.

\section{Injection of nonthermal pressures in clusters of galaxies by
radio galaxies}
\subsection{The jet power of radio galaxies}

\begin{figure}[t]
\setlength{\unitlength}{0.6\textwidth}
\psfig{figure=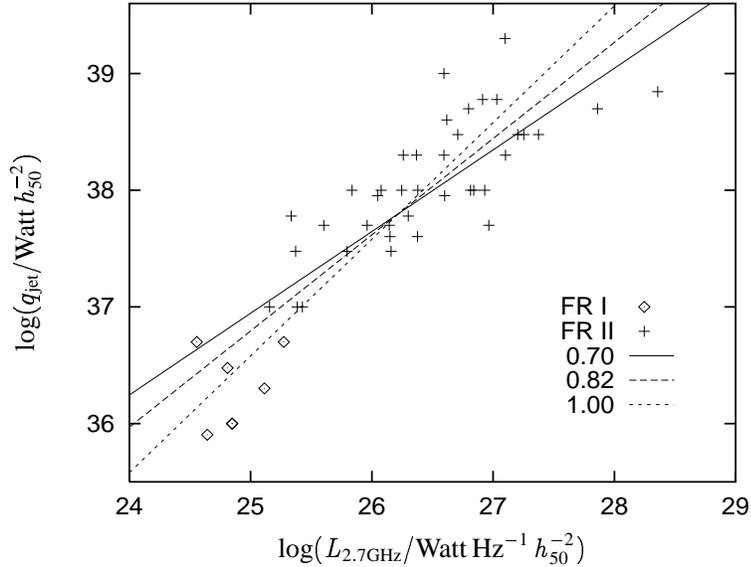,width=\unitlength}
\caption[]{\label{fig.Qjet.L27}Jet power of radio galaxies
$\log(q_{\rm jet}/{\rm Watt}\, h_{50}^{-2})$ from (Rawlings \&
Saunders 1991) \nocite{rawlings91} over their radio luminosity $\log(
L_{\rm 2.7GHz}/ {\rm Watt\,Hz^{-1}}\,h_{50}^{-2})$ from (Laing, Riley
\& Longair 1983) \nocite{laing83}. The power laws discussed in the
text $q_{\rm jet} \sim L_{\rm 2.7GHz}^{b}$ are shown.  Protons are
assumed in the jet corresponding to $f_{\rm power}=3$.}
\end{figure}

\placefigure{fig.Qjet.L27} 

%
%
Recent analysis of airshower data by Stanev
et al. (1995) \nocite{stanev95} and Hayashida et al. (1996)
\nocite{hayashida96} suggests a correlation between the arrival
directions of the highest energetic events of observed cosmic rays and
the supergalactic plane. This is consistent with earlier arguments
(Biermann \& Strittmatter 1987; Rachen \& Biermann 1993; Rachen,
Stanev, \& Biermann 1993) \nocite{biermann87,rachen93a,rachen93b} that
radio galaxies (RG) are sources of high energy cosmic rays. Of
interest for our present purposes is that these new observations imply
the production of very energetic relativistic protons, besides
relativistic electrons, in the lobes of radio galaxies.  In order to
explain the cosmic rays at energies beyond $3\cdot 10^{18}\,\mbox{eV}$
observed at Earth, we need to assume a high proton energy density within
lobes of radio galaxies. This requires that the power of the jet to be
up to an order of magnitude higher than the value which follows from
minimal energy arguments assuming no protons (Rachen \& Biermann 1993)
\nocite{rachen93a}. A factor $f_{\rm power}$ higher than 1 between
real jet power and the minimal possible jet power consistent with
radio observations can also be expected when we consider possible
acceleration mechanisms for protons, which can be very effective (Bell
1978a, 1978b) \nocite{bell78a,bell78b}.  { The recent gamma ray
detection of Mkn 421 (Petry et. al. 1996) has a natural explanation by
hadronic interactions, which implies that high energy protons are
indeed present within the jet-lobe system of RGs (Halzen 1996). } We
use here a conservative factor of $f_{\rm power} = 3$.  At this point
we link our argument for the importance of relativistic proton
energies to previous analyses of the energetics of extragalactic radio
source jets and lobes: In an interesting analysis, Rawlings \&
Saunders (1991)
\nocite{rawlings91} evaluated the jet power of a sample of radio
galaxies.  Beginning with the observed synchrotron emission of the
lobes Rawlings \& Saunders (1991) calculated the minimum necessary
power in relativistic electrons and magnetic fields which the jets
must inject into the radio lobes in order to produce the observed
radio emission. Since this calculation omitted the energy contribution
of the relativistic protons, we have increased these estimates of the
total jet power by $f_{\rm power} = 3$ times the value given by
Rawlings \& Saunders (1991). As indicated above, we feel that a factor
of 3 here is conservative. In a related investigation, Donea \&
Biermann (1996) \nocite{adonea96} have analysed recent AGN UV spectra
in terms of a sub-Eddington accretion disk which drives the innermost
parts of a radio galaxy jet. If this model is correct, then not only
relativistic protons but also a kinetic energy component of the
thermal gas exists $--$ which would drive $f_{\rm power}$ even higher.
Fig. \ref{fig.Qjet.L27} shows a double-logarithmic plot of the jet
power data $q_{\rm jet}$ (assuming $f_{\rm power}=3$) against $L_{\rm
2.7 GHz}$, the total radio luminosity at 2.7 GHz, derived from Laing,
Riley, \& Longair (1983) \nocite{laing83}. The relation is
\begin{equation}
q_{\rm jet} = 10_{}^{a} L_{\rm 2.7GHz}^{b}
\end{equation}
for which $b=0.82 \pm 0.07$, $a=38.28\pm 0.18 - 26.22 \cdot b$ ($cl =
68\%$), $q_{\rm jet}$ in Watt $h_{50}^{-2}$ and $L_{\rm 2.7GHz}$ in
Watt Hz$^{-1}\,h_{50}^{-2}$. This is consistent with the relation
derived by Falcke \& Biermann (1995) and Falcke, Malkan, \& Biermann
(1995) \nocite{falcke95a,falcke95b} which demonstrates that the radio
luminosity and jet power can be connected by a simple power law
through the entire spectrum of RG $--$ from the weaker, more
center-brightened FR I to the more powerful, edge-brightened FR II
galaxies. The same relation, which yields $b = 0.7$, has also been
successfully tested down to the scale of stellar jets and luminosities
(Falcke \& Biermann 1996) \nocite{falcke96}.
\subsection{The intracluster medium energy input from radio galaxies}
There is evidence for strong interaction between radio galaxies and
the ICM, since hot, compressed and very X-ray luminous regions are
observed within 20-200 kpc $h_{50}^{-1}$ from the radio galaxies in
clusters (Jones \& Forman 1984; B\"ohringer et al. 1993; Burns et
al. 1994; Sarazin, Baum \& O'Dea 1995)
\nocite{jones84,boeringer93,burns94,sarazin95}.  In this section, we
appeal to various statistical analyses which can be used to estimate
the likely extragalactic jet energy input into the ICM over a relevant
period of cosmic time.  We note, that ICM refers to the thermal gas,
the magnetic fields and the CRs and therefore energy input includes
heating of the gas, but also injection of CRs and fields.
The radio luminosity function (RLF) of entire clusters compared to
that of radio galaxies indicates that the fraction of the radio
galaxies which lie in clusters can be reasonably estimated to be
$f_{\rm cluster} \approx 0.3 ... 0.5$ (Owen 1975; Gubanov \&
Dagkesamanskii 1988)\nocite{owen75,gubanov88}. On a cosmic scale,
radio galaxies are significantly more clustered than galaxies in
general, or even elliptical galaxies (Bahcall \& Chokshi 1992)
\nocite{bahcall92}, and the bivariate luminosity function (optical \&
radio) of ellipticals inside and outside of clusters has been found to
be roughly the same, independent of cluster richness class and RG
radio power (Auriemma et al. 1977; Ledlow \& Owen
1995)\nocite{auriemma77,ledlow95}. In other words, the probability
that a potential progenitor of a jet-lobe radio galaxy will actually
produce one is not affected by the galaxy's membership in a
cluster. An additional, approximate scaling argument can be made as
follows: Cluster cores contribute $\approx $ 10 \% of the light in the
universe (Salpeter 1984) \nocite{salpeter84}, and radio galaxies in
rich clusters, especially the powerful ones, occur preferentially near
the center (Ledlow \& Owen 1995)\nocite{ledlow95}.  Given the
information on the similarity of the bivariant luminosity function
inside and outside of clusters, it could be supposed that the cores of
clusters contain 10 \% of all radio galaxies and therefore absorb some
fraction $f_{\rm core}$ of the total jet power.  We therfore
estimatete $f_{\rm core}\approx 0.1$.  The corresponding fraction
$f_{\rm cluster}$ absorbed by all clusters is $f_{\rm cluster}
\approx 0.3...0.5$.

In an attempt to arrive at some statistical estimate of the jet power
output of RG we have integrated the {\it evolving} part of the RLF
given in equation (7) in (Dunlop \& Peacock 1990) \nocite{dunlop90}.
Because of the sample chosen by Rawlings \& Saunders (1991)
\nocite{rawlings91} our $q_{\rm jet}$-$L_{\rm 2.7GHz}$ correlation is
assuredly valid above $10^{24.5}\,h_{50}^{-2} \mbox{W}\,
\mbox{Hz}^{-1}$, but it must be extrapolated to lower luminosities. To
accommodate uncertainties in this extrapolation, we calculated the
total jet power output, $Q_{\rm jet}(z)$, using three different power
law indices, $b=0.7,0.82$ and $1.0$ respectively.  The expression for
$Q_{\rm jet}$ is
\begin{equation}
Q_{\rm jet}(z) = \int \, \frac{dN(L_{\rm 2.7GHz}, z)}{dL_{\rm 2.7GHz}}
\, q_{\rm jet}(L_{\rm 2.7GHz}) \, dL_{\rm 2.7GHz}
\end{equation}

A numerical integration of the jet power-luminosity correlation gives
the average jet power delivered per cosmological comoving volume
$Q_{\rm jet}(z=0) = 11, 4, 2\,\cdot 10^{40}\,{\rm Watt \,
Gpc^{-3}}\,h_{50}^{-2}$, corresponding to the three power law slopes
mentioned above; $b= 0.70, 0.82, 1.00$ These are shown in
fig. \ref{fig.Qjet.L27}. The energy input into the central region of
clusters from RG is the fraction $f_{\rm core} \,Q_{\rm jet}(z=0)$,
which is $0.4 \cdot 10^{40}\,{\rm Watt \, Gpc^{-3}}\,h_{50}^{-2}$ for
$b=0.82$.  This is half an order of magnitude lower than the central
X-ray luminosity of clusters, about $10^{40}\,{\rm Watt \,
Gpc^{-3}}\,h_{50}$ which is, in turn, roughly $2/3$ of the total
luminosity for our $\beta$-model-parameters $r_{\rm cool}= 1.5
\,r_{\rm core}$ and $\beta =0.6$.
\subsection{Cooling of the cosmic ray protons vs accumulation}
In order to decide if the injected jet power dissipates and heats the
gas, or alternativly, accumulates and supports the ICM we must
estimate the time scale of the dissipation processes. The dissipation
of magnetic fields is very difficult to quantify, but the CR part of
the outflow of radio galaxies could be affectedby numerous processes.
The relativistic electrons lose their energy relatively quickly
through synchrotron emission in the cluster magnetic fields and
Compton scattering with photons of the microwave background. This
contrasts with the energetic protons, whose Compton and
synchrotron cooling times are much longer than the Hubble-time.
The energy loss of a proton with energy $\varepsilon = \gamma_{\rm p}
\, m_{\rm p} c^2$ by electronic excitations in a plasma is given by
Gould (1972)
\nocite{gould72}:
\begin{equation}
- \left( \frac{d\gamma_{\rm p}}{dt} \right)_{\rm ee} = \frac{4 \, \pi
 \, e^4 \,n_{\rm e}}{m_{\rm e} \, c^3 \, m_{\rm p} \, \beta_{\rm p} }
 \left[ \ln \left( \frac{2\gamma_{\rm p} m_{\rm e} c^2 \beta_{\rm
 p}^2}{\hbar
\omega_{\rm pl}} \right) - \frac{\beta_{\rm p} ^2}{2} \right]
\end{equation}
Here, $\beta_{\rm p}\, c$ is the velocity of the proton and
$\omega_{\rm pl} =
\sqrt{4 \pi e^2 n_{\rm e} /m_{\rm e}}$ the plasma frequency. Inserting
a typical density of $n_{\rm e} = 10^{-3} \, {\rm cm^{-3}}$ gives a
cooling time
\begin{equation}
t_{\rm ee} = \left[ - \frac{1}{\gamma_{\rm p}} \left(
\frac{d\gamma_{\rm p}}{dt} \right)_{\rm ee} \right]^{-1} = 4.7 \cdot
10^{10} \, \beta_{\rm p}\, \gamma_{\rm p} \,\left[ 1 +
\frac{\ln(\beta_{\rm p}^2 \,\gamma_{\rm p}) - \beta_{\rm p}^2 /2}{41} 
\right]^{-1} \, {\rm yrs}
\end{equation}
which is longer than a Hubble-time for $\beta_{\rm p} > 0.4$.
Cosmic rays streaming along a magnetic field are cooled by excitation
of Alfv\'{e}n waves (Wentzel 1974) \nocite{wentzel74}, which changes
their momenta $\beta_{\rm p}\gamma_{\rm p} m_{\rm p}$ 
\begin{equation}
- \left( \frac{d\beta_{\rm p}\gamma_{\rm p}}{dt} \right)_{\rm A} =
\frac{V_{\rm A} }{L}\, \beta_{\rm p} \gamma_{\rm p} 
\end{equation}
where $V_{\rm A}= B / \sqrt{4 \, \pi \, m_{\rm p} \, n_{\rm e}}$ is
the speed of the waves and $L$ is the scale height of the CR
distribution {  measured following } the magnetic
field. Rephaeli (1987) equates this length to the core radius $r_{\rm
core}$ of the cluster since the radial scale height of the gas and the
CR should be similar.  {  But $L = r_{\rm core}$ is correct only
in the case of radial magnetic fields in clusters, since then the CR
can move straight radially, and the length of a CR path is equal to the
Euclidean distance between starting and final point of the
path. }
In the more realistic case of tangled magnetic fields the path length
$L$ must be measured following the field lines, which maintain the
path for the CRs in order to arrive at a position one core radius in
radial distance from their starting point. This path-length must be of
the order $r_{core}^2/l_{\rm frs}$ due to the random walk of the field
line through the ICM, with a step size given by the field reversal
scale $l_{\rm frs}$. {  But the Euclidean scale height of the
CRs distribution, giving the pressure gradient which goes into the
hydrostatic equation, is still one core radius. } Magnetic fields
are typically $B \approx 2 \,\mu G (l_{\rm frs}/10 {\rm kpc})^{-1/2}$,
and a typical core radius is $r_{\rm core} = 250$ kpc. We get a
cooling time
\begin{equation}
t_{\rm A} = \frac{L}{V_{\rm A}} = 4.4\cdot 10^{10}\, \left(
\frac{l_{\rm frs}}{\rm 10 kpc} \right)^{-1/2}\, \left( \frac{n_{\rm
e}}{10^{-3}\,{\rm cm}^{-3}}\right)^{1/2}\, {\rm yrs}
\end{equation}
and of course a larger time if $l_{\rm frs}< 10$ kpc.
The time scale for energy loss by proton-proton interaction is of the
order (or even larger) of the average time between collisions of the
CR protons with the gas
\begin{equation}
t_{\rm pp} = ( \sigma_{\rm pp} \, n_{\rm e} \, \beta_{\rm p} \, c)^{-1} = 3.5
\cdot 10^{10} \, \beta_{\rm p}^{-1}\, \left( \frac{n_{\rm
e}}{10^{-3}\,{\rm cm}^{-3}}\right)^{-1}\,{\rm yrs}\, .
\end{equation}

There might be some adiabatic cooling of the CRs when they and their
enclosing magnetic fields ascend in the gravitational field of the
cluster. But the energy lost by the energetic particles is gained by
the magnetic fields and the thermal component of the ICM, and it
returns to the particles if the same plasma is dragged down by
infalling matter. Since clusters are still growing, V{\"o}lk,
Aharonian \& Breitschwerdt (1996) \nocite{voelk96} concluded that
there should be a sizeable adiabatic increase in the CR energy
content of the ICM.

Thus, the cooling of energetic protons is small, and may even be
compensated for by additional CR sources such as supernovae, accretion
shocks, in-situ acceleration and adiabatic compression.
Further, even the electrons can to some extent be re-accelerated by
energy redistribution in the ICM, as Kim et al. (1991) \nocite{kim91}
have suggested to explain the spectral index of the Coma cluster's
radio halo.

The number $f_{\rm power}$ depends on the ratio $k_{\rm p}= 
\varepsilon_{\rm p}/\varepsilon_{\rm e}$ between baryonic and electronic 
energy density within the radio lobes of radio galaxies. (In the case
of a lobe field energy density higher than the microwave energy
density: $f_{\rm power}= 1 + k_{\rm p}$). If one follows the radio
emitting medium which leaves the lobes and diffuses into the cluster
medium, this ratio increases strongly, because the electrons suffer
from numerous cooling mechanisms, where the protons nearly keep their
energy. This cooling of the electrons is clearly visible in the
steepening of the spectral index of radio emission close to radio
galaxies in clusters (e.g. Sarazin, Baum \& O'Dea 1995).  Therefore a
very high $k_p$ within the cluster medium is reasonable, indeed
necessary if one assumes the magnetic fields and CR energy density
values that we do.

\subsection{\label{eveff}Evolutionary effects of cosmological jet power}
\begin{figure}[tb]
\setlength{\unitlength}{0.6\textwidth}
\psfig{figure=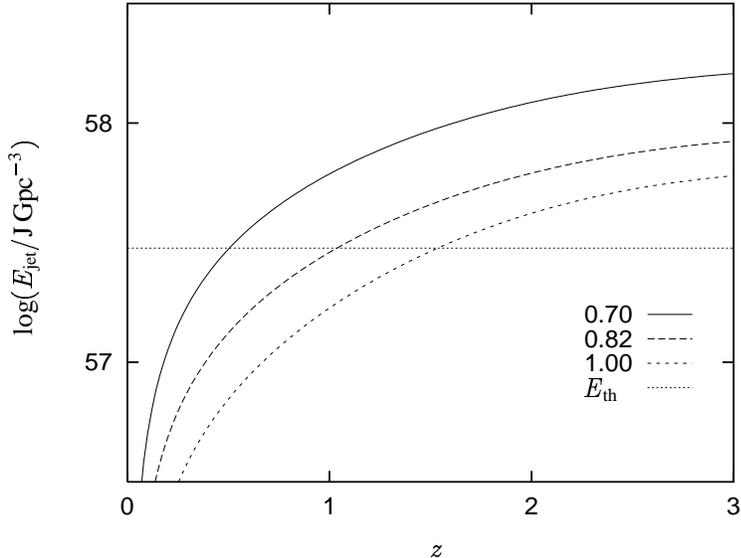,width=\unitlength}
\caption[]{\label{fig.Ecore}The time-integrated cosmological jet power
$\log(E_{\rm jet} / {\rm J \, Gpc^{-3}})$ which is injected into the
cores of cluster is shown as a function of the redshift z, which gives
the time of the lower end of the integration by equation
(\ref{eq.time}). The thermal Energy of the ICM in the cluster cores
$E_{\rm th}$ is shown.  ($f_{\rm power}=3$ and $f_{\rm core}=0.1$).}
\end{figure}

\placefigure{fig.Ecore}

Since the RLF is a strongly evolving function of the redshift z, the
energy input by RG must have been much stronger in the past. This is
of interest to calculate, since earlier jet input power might have
been {\it accumulated} in the thermal gas, cosmic rays, and magnetic
phase of the ICM.  Given a thermal energy content of the ICM within
the cooling, central region which is $E_{\rm th}(r_{\rm cool}) \approx
3\cdot 10^{57}\,h_{50}^{1/2}\, \mbox{J Gpc}^{-3}$ and $E_{\rm
th}(1.5\,{\rm Mpc})\, \approx 3\cdot 10^{58}\,h_{50}^{1/2}\,
\mbox{J Gpc}^{-3}$ then, as we have argued, the total accumulated
energy content should be of the order of $10^{58}\,h_{50}^{1/2}\,
\mbox{J Gpc}^{-3}$ within the central region, 
after allowing for cosmic rays and magnetic fields.

Evidence has been found for evolution of the cluster luminosity
function: Edge et al. (1990) \nocite{edge90} find that higher redshift
high-luminosity clusters are under-represented in their sample, which
fact they take to indicate strong evolution at recent epochs.
Castander et al. (1995) \nocite{castander95} confirm this observation,
and conclude that non-gravitational heating may be important in the
evolution of clusters. Further, a strong evolution of the luminosity
function in the Einstein 0.3 - 3.5keV band has been found over the
redshift range $z=0.17$ to $z=0.33$ (Henry et al. 1992)
\nocite{henry92} which is consistent with a lower ICM temperature at
higher redshift. This leads to the important consequence that the high
present day X-ray cooling of clusters cannot be extrapolated to the
past.

In fig. 2 we plot the accumulated jet power that was injected into
the cores of clusters (assuming $f_{\rm core}=0.1$) in the interval
since some past time (related to the redshift z) and the present,
using the standard expression for time in an Einstein-de Sitter
cosmology
\begin{equation}
\label{eq.time}
t(z) = \frac{2}{3\,H_o}\, \left[ 1- (1+z)^{-3/2} \right].
\end{equation}
We do not take adiabatic cooling due to the Hubble-expansion into
account because we are interested in the deposition of energy in
cluster cores, which are largely decoupled from the Hubble flow.  With
the above parameters of the $\beta$-model and of the efficiency of
injection of nonthermal energies ($f_{\rm power}$ \& $f_{\rm core}$)
we find that the thermal energy $E_{\rm th}$ of the ICM could have
been {\it accumulated} in less than a Hubble-time. The above mentioned
total energy content of $3\,E_{\rm th}$, could amount to a virtual
replacement of the thermal energy of the cooling region ICM by radio
galaxies and their ``fossils'', given the conservative level of our
relativistic proton energy component. At larger radii, the ratio of
nonthermal to thermal pressure should decrease, because the thermal
energy content is one order of magnitude higher within 1.5 Mpc than
within $r_{\rm cool}$ (for $\beta=0.6$) but the injected power only
increases by half an order of magnitude ($f_{\rm cluster}/f_{\rm core}
\approx 3 ... 5$). 

This is the key result of this paper: Using the best available
estimates of the power supplied by radio galaxies into the ICM, we
find that it balances (within 1 Mpc) or exceeds (within the core
region) the presently existing energy content of the gas including
magnetic fields and cosmic rays. An energy input of this order of
magnitude is unavoidable and must inevitably be present.

Additional CR energy sources such as supernovae, and shocks from
cluster merging events might also form a non-negligible part of the
energy budget. We have not attempted any quantitative estimate of the
latter two phenomena in this paper. The CR-density produced by
supernovae is calculated by V{\"o}lk, Aharonian \& Breitschwerdt
(1996) \nocite{voelk96} to be 3 - 30 \% of the thermal energy content
in the Perseus cluster.

\section{ Consequences for the cluster baryon fraction and cosmology}

We have demonstrated that radio galaxies can provide sufficient energy
to fill the central intracluster medium with magnetic fields and
cosmic ray protons which should be important within the 1 Mpc
scale. By analogy to the galactic disk, we surmise that an oversupply
will lead to instabilities (Parker 1969) \nocite{parker69} and hence
that {\it an approximate equipartition} between thermal, magnetic and
CR-pressure should result. We assume this to be true over a large
number of clusters. An additional nonthermal contribution to the
pressure could arise from turbulent motion of the ICM, which could
result from these instabilities.

The key result of our analysis above is that the X-ray - inferred gas
pressure $P_{\rm gas}$ is approximately a factor up to three lower
than the real total pressure $P_{\rm total}$. Thus a mass derived by
hydrostatics, 
\begin{equation}
- \frac{d P_{\rm total}}{dr} = \frac{G \, M(r) \, \varrho(r)}{r^2} \,
,
\end{equation}
where $M(r)$ is the mass within a radius of $r$ and $\varrho(r)$ is
the gas density derived from X-ray images, without consideration of
this nonthermal pressure should be up to a factor three less than the
correct mass within the cluster core. { The precise factor may vary
from cluster to cluster.} This would explain the reported mass
discrepancy between hydrostatic and lensing mass within the 1 Mpc
scale. It requires us to assume that virial mass estimates derived
from galaxy motions, which are on average as low as the hydrostatic
masses, are likewise often lower by a corresponding factor than the
correct masses.
The mass fraction of cluster gas $f_{\rm gas}=M_{\rm gas}/M_{\rm tot}$
measured at a few core radii will be obviously strongly affected by this
additional pressure.

Even the scatter in $f_{\rm gas}= (0.1 ... 0.3)\,h_{50}^{-3/2}$
derived by hydrostatics (Edge \& Stewart, G.~C.  1991, David, Jones \&
Forman 1995, Buote \& Canizares 1996), can be interpreted as
indicating that the limiting factor between total and thermal
pressures is $\sim 3$, and that this is reached in a number of
clusters. We assume that at least some of this scatter is due to
different values of the {\it additional} pressures in different
clusters and expect the true value of $f_{\rm gas}$ to be close to the
lower end of the range. In this context it is interesting to note that
Squires et al. (1996) \nocite{squires95}, who do not find a mass
discrepancy in the outer region of A2218, give a low value $f_{\rm
gas}= (0.11 \pm 0.06)h_{50}^{-3/2}$ for this region, which supports
our low baryon fraction.

At larger radii hydrostatic gas fractions become less reliable: Since
hydrostatic masses are usually derived with the asymptotics of a
$\beta$-model, they are extremely sensitive to the fitted value of
$\beta$. But Bartelmann \& Steinmetz (1996) show { using simulated
cluster data } that a cut-off radius in the radial X-ray profile
resulting from the X-ray background lowers the fitted value of $\beta$
significantly. If the `true' $\beta$ is one, 40 \% less thermal gas
and energy is within 1.5 Mpc, { and the possible range of
hydrostatical influence of RG increases}. The calculated total mass
increases with $\beta$ as Bartelmann \& Steinmetz (1996) note, and
therefore the gas fraction is strongly reduced at large scales, even
without sufficient nonthermal pressures. { Further, a
spectroscopically unresolved temperature decrease at large radii
(resolved by Markevitch (1996) and Markevitch, Sarazin \& Henrikson
(1996)) or strong decreasing nonthermal pressures would imply a
pressure gradient stronger than that of a $\beta$-model fitted with
X-ray data. The resulting total mass would be higher and the baryon
fraction would be lower than derived with the standard analysis. }

$f_{\rm gas}\approx 0.10 h_{50}^{-3/2}$ could be regarded as an upper
limit to the baryonic mass fraction $f_{\rm b} =\Omega_{\rm
b}/\Omega_o$ since it is not obvious that all the dark matter in the
universe is clumpy $f_{\rm gas} \geq f_{\rm b}$. The constraints from
nucleosynthesis are $0.04\,h_{50}^{-2}<\Omega_{\rm b}
<0.06\,h_{50}^{-2}$ (Walker et al. 1991) \nocite{walker91}
$...0.09\,h_{50}^{-2}$ (Copi, Schramm, \& Turner 1995) \nocite{copi95}
which adjusts the clumping part of the matter, which we shall identify
with $\Omega_{\rm cold}$, to
\begin{equation}
0.4 \, h_{50}^{-1/2}< \Omega_{\rm cold} < 0.6...0.9\, h_{50}^{-1/2}\,
.
\end{equation}
We note that $\Omega_{o}$ will be even higher than $\Omega_{\rm cold}$
if there is an appreciable mass fraction of hot dark matter outside of
rich clusters, in the universe.  Several recent papers (e.g. D.-W. Kim
\& Fabbiano 1995; White et al. 1993; Buote \& Canizares 1996)
\nocite{kim95,white93,buote96} have argued that there is a baryon
problem in clusters of galaxies, given that the data seem to show that
the baryonic mass fraction in clusters was higher than suggested by
nucleosynthesis for the entire universe.  This baryon problem
disappears in our approach, since the relative baryon mass is reduced
by a factor of up to 3.
\section{Discussion}
Starting with an analogy to the galactic disc, where nonthermal
pressures exceed the thermal pressure, we have demonstrated that a
similar nonthermal pressure component in clusters of galaxies could
close the gap between the mass derived from the observed scale hight
of the gas distribution and the mass derived from strong lensing,
which occurred in a number of clusters. One nonthermal phase would be
{ cluster core magnetic fields on the order of $10 \mu$G, directly
observed in some clusters with central radio galaxy and cooling
flow}, and probably typical of all clusters, if the (poorly known)
field reversal scale is as small as $\sim 0.1$ kpc.  Another phase
consists of cosmic ray protons, that have cooling times equal to or
larger than a Hubble-time. The most important source of fields and
energetic protons are probably radio galaxies, which are frequently
found in clusters. We derive an empirical law connecting the
synchrotron emission of the ``fast'' cooling electrons in the lobes of
radio galaxies and their jet power including electron, proton and
magnetic power. We integrate the evolution of the radio luminosity
function with this law and demonstrate that the injection of
nonthermal phases into the central cluster region during the lifetime
of a cluster should amount to several times the thermal content of the
ICM there.  Even within a one Mpc radius the injected energy should be
on the average comparable with the thermal content. { At larger
radii the average accumulated energy should be lower than the thermal
energy, unless radio galaxies are more powerful than we assumed or the
gas temperature decreases strongly at large radii (Markevitch 1996;
Markevitch, Sarazin \& Henrikson 1996). } We think it likely that some
fraction of the injected jet power is converted through instabilities
to turbulent ICM motion.

If this model is, as we propose, the principal solution to the mass
discrepancy (in combination with systematic errors at scales above 1
Mpc (e.g. Bartelmann \& Steinmetz 1996), a low baryon fraction $f_{\rm
b} \approx 0.1\,h_{50}^{-3/2}$ should result , which leads to a high
estimate of $\Omega_{\rm cold} \geq 0.4\, h_{50}^{-1/2}$.

\acknowledgements
This paper was conceived during a small meeting organized by the
Chinese Academy of Science and the Max Planck Society at Nandaihe,
Hebei, China, in August 1994.  PLB would like to acknowledge the
generous hospitality of his hosts at Beijing, Tianjin, Nanjing, Hefei
and Nandaihe. TAE would like to acknowledge the helpful discussions
with J{\"o}rg Rachen and Wolfram Kr{\"u}lls.



%
%
%


%
%


%
%
\end{document}